\documentclass[
showpacs,
floatfix,
aps,
prl,
amsmath,
twocolumn,
superscriptaddress,
tightenlines
groupaddress,
]{revtex4}

\usepackage{graphicx}
\usepackage{dcolumn}
\usepackage{amsmath}
\usepackage{amsfonts}
\usepackage{bm}
\usepackage{times}
\usepackage[varg]{txfonts}

\def\eph{\epsilon_{\mbox{\scriptsize ph}}}

\def\oc{\omega_{\mbox{\scriptsize {c}}}}

\def\os{\omega_s}

\def\vh{v_{\mbox{\scriptsize {H}}}}
\def\tq{\tau_{\mbox{\scriptsize {q}}}}

\def\tph{\tau_{\mbox{\scriptsize {ph}}}}
\def\muph{\mu^{\mbox{\scriptsize {ph}}}}
\def\muim{\mu^{\mbox{\scriptsize {im}}}}
\def\ttrim{\tau_{\mbox{\scriptsize {tr}}}^{\mbox{\scriptsize {im}}}}

\def\tqim{\tau_{\mbox{\scriptsize {q}}}^{\mbox{\scriptsize {im}}}}

\def\vtpar{v^\parallel_{\mbox{\scriptsize {t}}}}

\def\tqee{\tau_{\mbox{\scriptsize {q}}}^{\mbox{\scriptsize {ee}}}}

\bibpunct{[}{]}{,\!}{n}{,}{,} %prl
\begin{document}

\title{
Phonon-induced Resistance Oscillations in Very-high Mobility 2D Electron Systems
}

\author{A.\,T. Hatke}
\author{M.\,A. Zudov}
\email[Corresponding author: ]{zudov@physics.umn.edu}
\affiliation{School of Physics and Astronomy, University of Minnesota, Minneapolis, Minnesota 55455, USA} 
\author{L.\,N. Pfeiffer}
\author{K.\,W. West}
\affiliation{Bell Labs, Alcatel-Lucent, Murray Hill, New Jersey 07974, USA}
%\received{November 2, 2008}
%; revised manuscript received }

\begin{abstract}
We report on temperature dependence of acoustic phonon-induced resistance oscillations in very-\-high mobility two-dimensional electron systems. 
We observe that the temperature dependence is non-monotonic and that higher order oscillations are best developed at progressively lower temperatures.
Our analysis shows that, in contrast to Shubnikov-de Haas effect, phonon-induced resistance oscillations are sensitive to electron-electron interactions modifying the single particle lifetime.

\end{abstract} 
\maketitle

When a two-dimensional electron system (2DES) is subject to a weak perpendicular magnetic field $B$ and low temperature $T$, the linear response resistivity exhibits well-known Shubnikov-de Haas oscillations (SdHO) \citep{shubnikov:1930}.
SdHO are controlled by the ratio of the Fermi energy, $E_F$, to the cyclotron energy, $\hbar\oc$, and are periodic in $1/B$.
From the magnetic field and temperature dependences of the oscillation amplitude one can deduce quantum scattering time $\tau_q$ and the effective mass $m^*$ of the charge carrier.

Over the past decade it was realized that other types of low-$B$ resistance oscillations can be ``induced'' in 2DES by microwave \citep{miro} or dc \citep{hiro:exp} electric fields (or their combination \citep{miro:dc:exp}).
These oscillations originate from inter-Landau level transitions owing to microwave absorption and/or scattering off of disorder, respectively.
Microwave-induced oscillations are governed by $\omega/\oc$ ($\omega=2\pi f$ is the microwave frequency) and dc-induced oscillations are controlled by $2 k_F \vh/\oc$ ($k_F$ is Fermi wave number, $\vh$ is the Hall drift velocity).
As a result, $m^*$ is available directly from the oscillation frequency.
Microwave-induced oscillations have been extensively studied both theoretically \citep{miro:th} and experimentally \citep{miro:exp}, in part, because of their ability to evolve into zero-resistance states \citep{zrs}.
Recently, it was realized that dc fields can induce similar states with zero-differential resistance \citep{bykov:2007,zhang:2008} and theoretical proposals have been put forward \citep{hiro:th}.

Another class of oscillations in 2DES emerges at elevated temperatures when acoustic phonon modes with Fermi momentum become populated \citep{zudov:2001b}. 
These oscillations are called phonon-induced resistance oscillations (PIRO).
Remarkably, even though phonons of many different energies are present, the main contribution comes from the most energetic phonons an electron can scatter off.
This corresponds to a momentum transfer $\simeq 2k_F$, provided by an acoustic phonon.
The momentum conservation then selects a frequency $\os=2 k_F s$ from the phonon dispersion which is determined only by $k_F$ and the sound velocity $s$.
The energy of this phonon is used by an electron to make indirect transitions between Landau levels and the resistance oscillates with $\eph\equiv 2k_F s/\oc$.

There exist several issues that warrant investigations of PIRO.
First, experiments in GaAs-based structures reported a variety of sound velocities ranging from $3$ km/s \citep{zudov:2001b,zhang:2004} to $6$ km/s \citep{bykov:2005c,zhang:2008}.
Second, it is still unclear whether PIRO are caused by {\em interface} or {\em bulk} phonons \citep{piro:th}.
Finally, little is known about the contribution from different phonon modes and the temperature dependence.

In this Letter we study PIRO in a very high mobility 2DES over a wide temperature range.
In contrast to previous studies, we find that PIRO persist down to temperatures $\lesssim 2$ K (well below $\hbar\os/k_B \simeq 8 $ K) and extend to much lower magnetic fields exhibiting up to eight oscillations.
Such remarkable quality allows us to study how PIRO evolve with temperature.
We find that {\em higher} $B$ oscillations fully develop at progressively {\em higher} temperatures and that this ``optimal'' temperature roughly scales with $\sqrt{B}$. 
This observation suggests that PIRO, unlike SdHO, decay at higher temperature due to electron-electron interactions modifying the quantum lifetime entering the {\em square} of the Dingle factor.

Our Hall bar sample (width $w=100$ $\mu$m) was fabricated from a symmetrically doped GaAs/Al$_{0.24}$Ga$_{0.76}$As 300-\AA-wide quantum well grown by molecular-beam epitaxy.
The impurity-limited mobility $\muim$ and the density $n_e$ were $\simeq 1.2 \times 10^7$ cm$^2$/Vs and $3.75 \times 10^{11}$ cm$^{-2}$, respectively. 
The experiment was performed at temperatures from $2$ K to $7$ K.
Resistivity $\rho$ was measured using standard low-frequency (a few Hz) lock-in technique in sweeping magnetic field.

%\section{Fig. 1}
%%%%%%%%%%%%%%%%%%%%%%%%%%%%%%%%%%%%%%%%%%%%%%%%%
%fig 1
\begin{figure}[t]
%\resizebox{0.5\textwidth}{!}{
\includegraphics{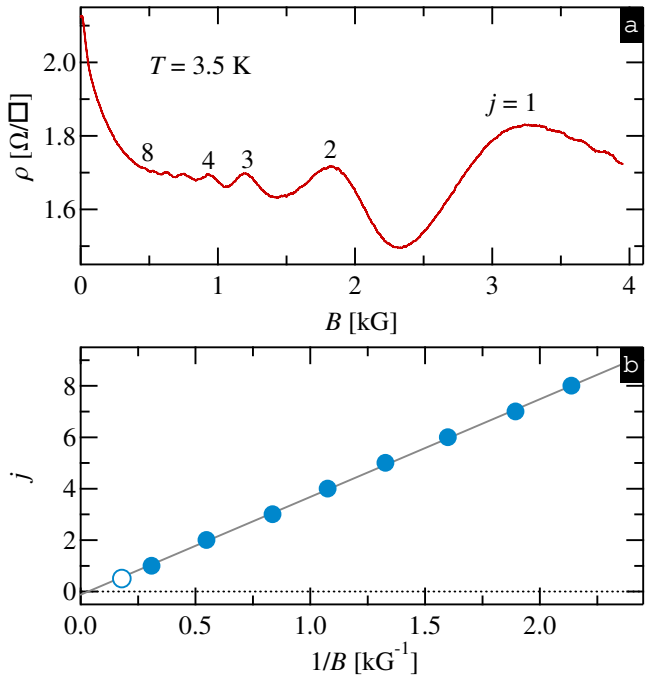}
%}
%\vspace{-0.1 in}
\caption{[color online]
(a) Magnetoresistivity $\rho$ at $T = 3.5\,$K showing up to eight PIRO peaks marked by integers.
(b) PIRO order $j$ vs. $1/B$ [filled circles]. 
Linear fit yields $s\simeq 3.4\,$km/s.
Open circle marks the extra peak [cf.\,$\downarrow$ in Fig.\,2].
}
%\vspace{-0.15 in}
\label{f1}
\end{figure}
%%%%%%%%%%%%%%%%%%%%%%%%%%%%%%%%%%%%%%%%%%%%%%%%%
%(a)
In Fig.\,\ref{f1}\,(a) we present resistivity $\rho(B)$ measured at $T = 3.5\,$K.
At this temperature, SdHO are strongly suppressed, but PIRO \citep{note} exhibit up to eight oscillations, as marked by integer $j$.
This confirms excellent quality of PIRO in our sample since in previous studies even the third oscillation was barely resolved.
We further find that PIRO persist down to $B\simeq 500$\,G, about one order of magnitude lower than the onset of the SdHO.
%(a)
To extract the sound velocity of the relevant phonon mode we present in Fig.\,\ref{f1}(b) the PIRO order $j$ as a function of $1/B$.
The data roughly fall onto a line $j=(2k_F)s/\oc \propto 1/B$ and the fit reveals sound velocity $s\simeq 3.4$\,km/s, corresponding to $\hbar\os\simeq 8$ K.
We note that this value is close to the velocity of the bulk transverse acoustic phonons, $\vtpar=3.35$ km/s. 

%\section{Fig. 2}
%%%%%%%%%%%%%%%%%%%%%%%%%%%%%%%%%%%%%%%%%%%%%%%%%
%fig 2
\begin{figure}[t]
%\resizebox{0.5\textwidth}{!}{
\includegraphics{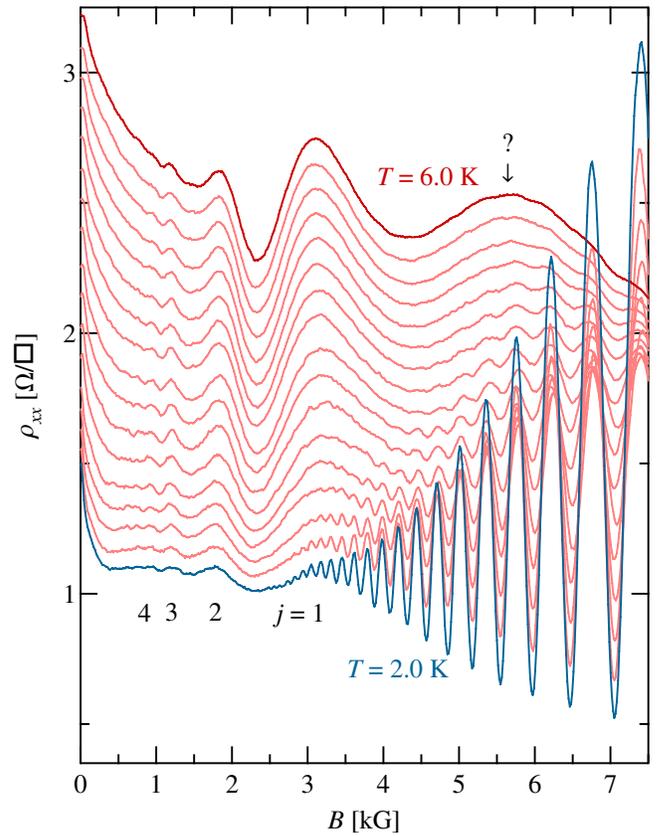}
%}
%\vspace{-0.1 in}
\caption{[color online]
Magnetoresistivity $\rho(B)$ at different temperatures, from $2.0\,$K (bottom trace) to $6.0\,$K (top trace), in $0.25\,$K increments.
Traces are {\em not} vertically offset. 
Integers at the bottom trace mark the order $j$ of the PIRO peaks.
}
%\vspace{-0.15 in}
\label{f2}
\end{figure}
%%%%%%%%%%%%%%%%%%%%%%%%%%%%%%%%%%%%%%%%%%%%%%%%%
%fig2
We now turn to the temperature evolution of PIRO. 
In Fig.\,\ref{f2} we show $\rho(B)$ at temperatures from $2\,$K to $6\,$K in $0.25\,$K increments.
In previous studies employing lower-mobility samples, PIRO were observed at relatively high temperatures, of the order of $\hbar\os$ (typically from $5$ to $20$ K), and SdHO were virtually absent. 
In contrast, our high mobility sample reveals {\em coexisting} PIRO and SdHO at lower temperature (cf., bottom trace at $T=2.0$ K). 
With increasing temperature, SdHO uniformly decay, as expected, while PIRO show more intriguing behavior;
PIRO initially grow, then saturate, and eventually weaken. 
More importantly, the optimal temperature ({\em i.e.,} at which a particular peak is best developed) depends on the order $j$.
As we will show, lower order (higher magnetic field) peaks are best developed at progressively higher temperature than the higher order (lower magnetic field) peaks. 

In addition to the PIRO series shown in Fig.\,\ref{f1}\,(a), higher temperature data reveal a distinct peak emerging near $B\simeq 5.7$ kG (cf.,\,$\downarrow$).
One possible scenario is that this peak originates from another, faster phonon mode which gets populated at higher temperatures.
From $\eph=1$ we estimate sound velocity of this mode as 4.8 km/s which compares favorably with the velocity of the bulk longitudinal acoustic phonons in GaAs.
Another possible origin is a two-phonon process (or two consecutive single-phonon processes), similar to that responsible for fractional microwave-induced oscillations.
Assuming the value of the sound velocity extracted earlier, $s\simeq 3.4$ km/s, and associating the peak with $\eph \simeq 1/2$, we add the extra point to Fig.\,\ref{f1}(b) [open circle] and observe reasonable agreement with the rest of the data.

%\section{Fig. 3}
%%%%%%%%%%%%%%%%%%%%%%%%%%%%%%%%%%%%%%%%%%%%%%%%%
%fig 3
\begin{figure}[t]
%\resizebox{0.5\textwidth}{!}{
\includegraphics{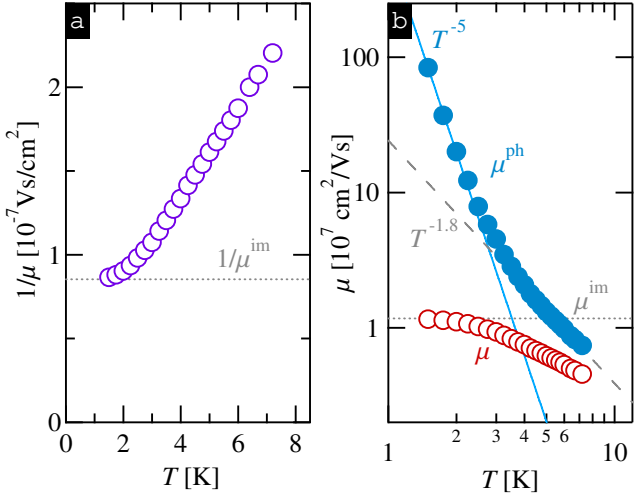}
%}
%\vspace{-0.1 in}
\caption{[color online]
(a) Inverse mobility $1/\mu$ vs. $T$.
(b) Total mobility $\mu$ (open circles), impurity mobility $\muim$ (dotted line) and acoustic-phonon mobility $\muph$ (filled circles) vs. $T$. 
}
%\vspace{-0.15 in}
\label{f3}
\end{figure}
%%%%%%%%%%%%%%%%%%%%%%%%%%%%%%%%%%%%%%%%%%%%%%%%%
To further analyze our results we examine the temperature dependence of the electron mobility.
In Fig.\,\ref{f3}\,(a) we present $1/\mu=e n_e\rho(0)$ as a function of temperature and observe significant increase.
This behavior reflects the crossover to the Bloch-Gr\"uneisen regime occurring at $k_B T \simeq \hbar\os$ \citep{bg}.
Assuming Matthiessen's rule, the total mobility is $1/\mu (T)=1/\muim+1/\muph (T)$, where $1/\muim$ and $1/\muph$ are impurity and acoustic-phonon contributions, respectively.
At low temperature, acoustic-phonon contribution is negligible and $\mu \simeq \muim$, as demonstrated in Fig.\,\ref{f3}\,(a).
Following the established procedure for data reduction \citep{bg}, we extract $\muph$ and $\muim$ and present their temperature dependencies in Fig.\,\ref{f3}\,(b) on a log-log scale.
Here, $\muph$ [full circles] was deduced from the total mobility [open circles] assuming $\muim=1.17\times 10^7$ cm$^2/$Vs [dotted line].
For low temperatures, we observe $\muph \propto T^{-5}$ [solid line], consistent with the earlier experiments in similar mobility 2DES \citep{bg}.
At higher temperatures, the dependence slows down and at $k_B T > \hbar\os$ should approach $\muph \propto 1/T$ \citep{price}.
Fitting with the power law $\muph \propto T^\alpha$ over the experimentally relevant temperature range yields $\alpha \simeq 1.8$ [dashed line].

\begin{figure}[t]
%\resizebox{0.5\textwidth}{!}{
\includegraphics{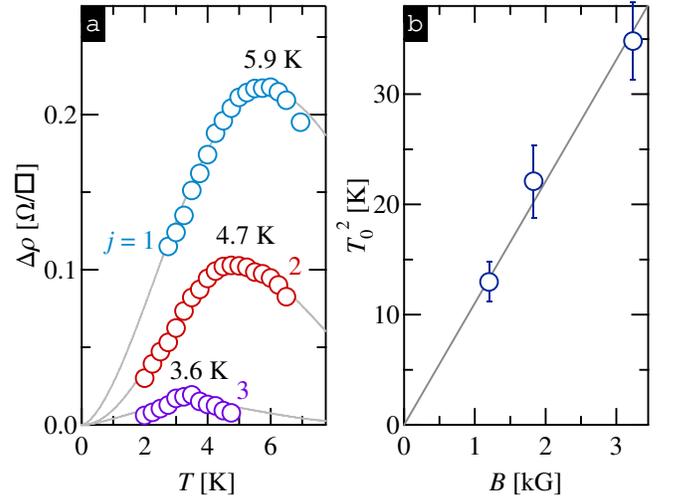}
%}
%\vspace{-0.1 in}
\caption{[color online]
(a) PIRO amplitude at $j=1$, $2$, and $3$ vs. $T$. 
Curves are calculated using Eq.\,\ref{th} for $\lambda=4.1$ and $\alpha=1.8\pm0.2$.
(b) Optimal temperature $T_0^2$ vs. $B$.
}
%\vspace{-0.15 in}
\label{f4}
\end{figure}

We now extract the PIRO amplitude for the first three peaks, $j=1,2,3$ and present the results in Fig\,\ref{f4}\,(a) as a function of temperature.
All three data sets exhibit non-monotonic behavior showing a maximum at the optimal temperature $T_0$, which increases with magnetic field. 
One of the factors contributing to the decay of the PIRO amplitude at low temperatures is the decreasing population of the $2k_F$ phonons required for phonon absorption.
At the same time, phonon emission is also suppressed at low temperatures as the electrons become more degenerate and less capable of emitting $2k_F$ phonons.
Thus both the emission and the absorption of phonons is suppressed by Fermi and Plank distributions, respectively. 

To explain the decay at higher temperatures, we recall that PIRO originate from inter-Landau level transitions and thus rely on both initial, $\nu(\varepsilon)$, and final, $\nu(\varepsilon+\hbar\os)$, densities of states.
In the regime of overlapped Landau levels, $\nu(\varepsilon) = \nu_0[1-2\delta \cos(2\pi \varepsilon/\hbar\oc)]$, where $\delta=\exp(-\pi/\oc\tq)$ is the Dingle factor and $\nu_0=\nu(B=0)$.
The leading $\eph$-dependent contribution to the resistivity originates from the $\delta^2$ term generated by the product of the oscillatory parts of the corresponding densities of states.
This term survives averaging over the Fermi distribution, $\left < \cos^2(2\pi\varepsilon/\oc) \right >_\varepsilon \simeq 1/2$, and therefore, unlike SdHO, PIRO are insensitive to the temperature smearing of the Fermi surface.
This fact is manifested in Fig.\,2 by a much slower decay of PIRO with increasing temperature, as compared to SdHO.  

We further recall that in clean 2DES electron-electron interactions can significantly modify quantum scattering rate $1/\tq$  \citep{ts:ee,miso:ee}.
Taking the standard estimate for the electron-electron scattering rate \citep{chaplik:giuliani}, $1/\tqee=\lambda T^2/E_F$, where $\lambda \sim 1$, we find that $1/\tqee$ becomes comparable to impurity contribution $1/\tqim$ at just a few Kelvin in our 2DES (we estimate $\tqim \simeq 15$ ps). 
In principle, one can also consider electron-phonon scattering.
At temperatures of interest, electron-phonon scattering is not confined to small angles and therefore its contribution, $1/\tph$, can be estimated from $\muph$.
However, as shown in Fig.\,3(b), $1/\tph \lesssim 1/\ttrim\ll 1/\tq$ and thus electron-phonon contribution to the quantum scattering rate can be safely ignored. 
%\section{Fig. 4}
%%%%%%%%%%%%%%%%%%%%%%%%%%%%%%%%%%%%%%%%%%%%%%%%%
%fig 3
%%%%%%%%%%%%%%%%%%%%%%%%%%%%%%%%%%%%%%%%%%%%%%%%%

Assuming $1/\tq=1/\tqim+1/\tqee$, we write the temperature-dependent part of the PIRO amplitude:
\begin{equation}
\Delta \rho(T) \propto \tph^{-1}(T) \exp[-2\pi/\oc\tqee(T)].
\label{th}
\end{equation}
Here, $\tph^{-1}(T)$ is responsible for the initial growth of PIRO and $\exp(-2\pi/\oc\tqee)$ accounts for the decay at higher temperatures.
As a result, PIRO amplitude exhibits a maximum at some optimal temperature $T_0$ which increases with the magnetic field.
This is exactly what we observe in Fig.\,\ref{f4}(a).
Assuming $1/\tph \propto T^\alpha$, we estimate the optimal temperature as $k_BT_0 = (\alpha E_F\hbar\oc/4\pi\lambda)^{1/2} \propto \sqrt B$.
In Fig.\,\ref{f4}(b) we plot $T_0^2$ as a function of magnetic field and observe roughly linear dependence with the slope corresponding to $\lambda \simeq 4.1$.
This suggests that $\tqee \simeq \tqim$ at $T\simeq 4.5$ K.

In summary, we have studied phonon-induced resistance oscillations in a very-high mobility 2DES.
Owing to low disorder in our samples oscillations persist down to very low temperatures (and very low magnetic fields) allowing for detailed studies of temperature dependence. 
We have found that the temperature dependence reveals a pronounced maximum marking the optimal temperature at which a particular oscillation is best developed. 
This optimal temperature increases with the magnetic field as $\sqrt{B}$ which we attribute to the electron-electron interaction effects modifying quantum scattering rate entering the Dingle factor.
We note that such contribution from electron-electron interactions is not available from conventional magneto-transport, such as SdHO \citep{martin:adamov}.

We thank I. Dmitriev, M. Vavilov and B. Shklovskii for discussions and W. Zhang for assistance with initial experiments.
This work was supported by NSF Grant No. DMR-0548014.
\vspace{-0.25in}
%\bibliography{../../bibliography_final_2}

\begin{thebibliography}{21}
\bibitem{note}
{Oscillation frequency was checked to scale with $\sqrt{n_e}$ ($n_e$ was tuned by persistent photoconductivity effect).}
\end{thebibliography}

\end{document}